\documentclass[onecolumn,
amsmath,amssymb 
]{revtex4}

\usepackage{bm}
\usepackage{graphicx}
\usepackage{amsbsy} 
\usepackage{amsmath}
\usepackage{amsfonts}
\usepackage{amsthm}
\usepackage{bm}
\usepackage{graphicx}
\usepackage{amsbsy}
\usepackage{amsmath}
\usepackage{amsfonts}
\usepackage{amsthm}
\usepackage{braket}
\usepackage{color}
\usepackage{mathrsfs}
\usepackage{changes}

\begin{document}

\theoremstyle{plain}
\newtheorem{theorem}{Theorem}
\newtheorem{lemma}[theorem]{Lemma}
\newtheorem{corollary}[theorem]{Corollary}
\newtheorem{conjecture}[theorem]{Conjecture}
\newtheorem{proposition}[theorem]{Proposition}

\theoremstyle{definition}
\newtheorem{definition}{Definition}

\theoremstyle{remark}
\newtheorem*{remark}{Remark}
\newtheorem{example}{Example}

\title{Quantum cryptographic resource distillation and entanglement}

\author{Minjin Choi}
\affiliation{Department of Mathematics and Research Institute for Basic Sciences, Kyung Hee University, Seoul 02447, Korea} 

\author{Soojoon Lee}
\affiliation{Department of Mathematics and Research Institute for Basic Sciences, Kyung Hee University, Seoul 02447, Korea}

\date{\today}

\begin{abstract}
We look into multipartite quantum states 
on which quantum cryptographic protocols including quantum key distribution and quantum secret sharing 
can be perfectly performed, 
and define the quantum cryptographic resource distillable rate 
as the asymptotic rate at which such multipartite state can be distilled from a given multipartite state. 
Investigating several relations between entanglement and the rate, 
we show that there exists a multipartite bound entangled state whose quantum cryptographic resource distillable rate is strictly positive, 
that is, there exists a multipartite entangled state which is not distillable, 
but can be useful for quantum cryptography such as quantum key distribution and quantum secret sharing. 
\end{abstract}

\maketitle

\section{Introduction}

Entanglement is one of the most significant resources for quantum cryptography. 
In particular, it has been well known that 
any pure entangled state can be useful in performing quantum cryptographic protocols, 
such as quantum key distribution~\cite{E91} and quantum secret sharing~\cite{HBB99}.  
However, it has also been known that
there exist mixed states, called the private states~\cite{HO05,HA06,H09} or the (genuine) secret sharing states~\cite{CL08,CL21},
which can distill perfectly secure key bits or secret bits for secret sharing just by measurement.
We here call such mixed states the {\em quantum cryptographic resource} (QCR) states. 
Hence, it can be seen that 
a QCR state is not only considered as a generalized version of the private state or the genuine secret sharing state, but is also
regarded as a resource unit in a quantum cryptographic theory, 
while a pure maximally entangled state plays a role of a resource unit in entanglement theory.  

We consider a general form of the QCR states with one dealer party.
In other words, the QCR state that we here deal with is a multipartite quantum state, 
and a private state on the parties can be obtained from the state 
by local quantum operations and classical communication (LOCC)
so that perfectly secure key distribution is feasible between the dealer party and any player party of the state.
In addition, complete secret sharing on any number of divided parties 
together with the dealer party of the QCR state is also possible, 
although dishonest players cooperating with any exterior eavesdropper exist.   
Thus, players can select one quantum cryptographic protocol among various kinds of ones with the dealer on the same QCR state, 
as they want. 
 
As in any resource theories including entanglement theory,
it is both natural and important  
to take into account the quantity 
representing how much amount of QCR can be extracted from a given state, 
which we call the {\em QCR distillable rate} of the state. 
We remark that since the simplest QCR state is a maximally entangled state, 
the QCR distillable rate in entanglement theory is nothing but the distillable entanglement~\cite{BDSW96},
and since the private state is also a simple form of the QCR state, 
the QCR distillable rate in quantum key distribution is equal to the distillable key rate~\cite{HO05,H09,CL07}.
Hence, in this paper, we discover the properties of the QCR distillable rate, 
and compare the QCR distillable rate with the distillable entanglement and the distillable key rate.

We say that a multipartite state is {\em QCR distillable} 
if its QCR distillable rate is strictly positive. 
Then it is clear that 
a QCR distillable state is entangled, 
since if a multipartite state has a separable bipartite split, 
then perfectly secure key distribution is impossible between the split, 
and hence the state is not QCR distillable. 
However, it does not seem to be true that all entangled states are QCR distillable, 
because its simplest case is not true, that is, 
there exists a bipartite bound entangled state with positive secret key distillable rate~\cite{HO05,H09,CL07}.

In this paper, we first present necessary and sufficient conditions for the QCR state with a dealer party, 
and definition of the QCR state, 
and then show that a given multipartite quantum state is a QCR state 
if and only if the conditions on the state hold. 
We also define the QCR distillable rate of a given multipartite state in a mathematical way, 
and present some properties on the QCR distillable rate. 
Finally, by providing the method to construct a QCR state with larger number of parties from several QCR states, 
we prove the existence of multipartite bound entangled but QCR distillable states. 

This paper is organized as follows.
In Sec.~\ref{Sec:QCR}, we first define the QCR state, and justify the definition. 
After showing the several properties of the QCR states, 
we also present the mathematical definition of the QCR distillable rate of a given state in Sec.~\ref{Sec:QCR_rate}, 
and investigate some relations between the QCR distillable rate and 
other distillable rates such as the entanglement distillable rate and secret key distillable rate. 
We finally show that there exists a multipartite QCR distillable state without any distillable entanglement. 


\section{QCR states}\label{Sec:QCR}

Assume that there are one dealer and $N$ players 
who participate in a quantum cryptographic protocol,
and let $D=\bar{D}\widetilde{D}$ be the dealer's quantum system 
with two subsystems $\bar{D}$ of $d$ dimension and $\widetilde{D}$ of arbitrary dimension
in the protocol. 
Similarly, for each $1 \le i \le N$,
let $A_{i}=\bar{A}_i\widetilde{A}_i$ be the $i$-th player's quantum system 
with subsystems $\bar{A}_i$ of $d$ dimension and $\widetilde{A}_i$ of arbitrary dimension.
Throughout this paper, we denote $\bar{A}_{1} \cdots \bar{A}_{N}$ and $\widetilde{A}_{1} \cdots \widetilde{A}_{N}$ 
by $\mathbf{\bar{A}}$ and $\widetilde{\mathbf{A}}$, respectively, 
and denote $\mathbf{\bar{A}}\widetilde{\mathbf{A}}$ by $\mathbf{A}$.
The systems $\bar{\mathbf{A}}$ and $\mathbf{\widetilde{A}}$ are called the {\em information part} and the {\em shield part}, respectively.

In order to perform the quantum cryptographic protocol,
the dealer's and players' information should satisfy 
the following cryptographic conditions:
\begin{enumerate}
\label{CR_conditions} 
\renewcommand{\theenumi}{\roman{enumi}}
\renewcommand\labelenumi{(\theenumi)}
\item\label{condition1} 
The probability distributions of the dealer's and all players' information
must be unbiased and perfectly correlated.
\item\label{condition2}
An eavesdropper and dishonest players cannot 
get any information about the dealer's information.
\item\label{condition3}
The dealer and any subset of players can perform the same protocol with smaller number of parties after properly  applying LOCC.
\end{enumerate}
When $N=1$, if the dealer and the player share the private states~\cite{HO05,HA06,H09}, or 
when $N\ge 2$, if the dealer and the players share the genuine secret sharing states~\cite{CL21},
then the above three conditions are surely satisfied.
However, since any player can be a dealer 
in the private states and the (genuine) secret sharing states, 
considering the case where  the dealer is predetermined 
is more general than those in the private states and the (genuine) secret sharing states, 
Thus, we introduce the class of quantum states
suitable for the case where the dealer is determined in advance.

\begin{definition}
\label{Def:QCR}
$\Upsilon_{D\mathbf{A}}$ is called a {\em QCR state} 
if for any bipartite split $\{\mathbf{P}_1, \mathbf{P}_2\}$ of the players 
with $\mathbf{P}_1$ consisting of at least one player and $\mathbf{A}= \mathbf{P}_1\mathbf{P}_2$,
the given state $\Upsilon_{D\mathbf{A}}$ can be written as
\begin{align}
\label{QCR_state}
\frac{1}{d^{N}}
\sum_{iI_{1}I_{2}, jJ_{1}J_{2} \in \mathfrak{S}_{N}^{0}}
\ket{iI_{1}I_{2}}_{\bar{D}\mathbf{\bar{P}}_1\mathbf{\bar{P}}_2}\bra{jJ_{1}J_{2}}
\otimes
\left(U_{\widetilde{D}\mathbf{\widetilde{P}}_{1}}^{iI_{1}}V^{{I}_{2}}_{\widetilde{D}\mathbf{\widetilde{A}}}\right)
\sigma_{\widetilde{D}\mathbf{\widetilde{A}}}
\left(U_{\widetilde{D}\mathbf{\widetilde{P}}_{1}}^{jJ_{1}}V^{J_{2}}_{\widetilde{D}\mathbf{\widetilde{A}}}\right)^{\dagger},
\end{align}
where 
\begin{equation}
\mathfrak{S}_{N}^{t} \equiv 
\left\{jJ=jj_{1}j_{2} \cdots j_{N} \in \mathbb{Z}_{d}^{N+1}: 
j+\sum_{k=1}^{N}j_{k}\equiv t \pmod{d} \right\},
\label{eq:S_N}
\end{equation}
$\bar{D}\mathbf{\bar{P}}_1\mathbf{\bar{P}}_2=\bar{D}\mathbf{\bar{A}}$  
and $\widetilde{D}\mathbf{\widetilde{A}}$
are the information part and the shield part of the QCR state $\Upsilon_{D\mathbf{A}}$, respectively,
$\sigma_{\widetilde{D}\mathbf{\widetilde{A}}}$ is an arbitrary state,
and the $\left\{U_{\widetilde{D}\mathbf{\widetilde{P}}_{1}}^{iI_{1}}\right\}$ 
and $\left\{V^{{I}_{2}}_{\widetilde{D}\mathbf{\widetilde{A}}}\right\}$
are unitary operators on the systems $\widetilde{D}\mathbf{\widetilde{P}}_1$ and $\widetilde{D}\mathbf{\widetilde{A}}$, respectively.
\end{definition}

For instance, let $\ket{\Upsilon}_{\bar{D}\bar{A}\bar{B}\widetilde{D}\widetilde{A}\widetilde{B}}$ be the following state.
\begin{align}
\ket{\Upsilon}_{\bar{D}\bar{A}\bar{B}\widetilde{D}\widetilde{A}\widetilde{B}}=
\frac{1}{2}
&(\ket{000}_{\bar{D}\bar{A}\bar{B}}\ket{000}_{\widetilde{D}\widetilde{A}\widetilde{B}}
+\ket{011}_{\bar{D}\bar{A}\bar{B}}\ket{100}_{\widetilde{D}\widetilde{A}\widetilde{B}} 
+\ket{101}_{\bar{D}\bar{A}\bar{B}}\ket{100}_{\widetilde{D}\widetilde{A}\widetilde{B}}
+\ket{110}_{\bar{D}\bar{A}\bar{B}}\ket{000}_{\widetilde{D}\widetilde{A}\widetilde{B}}). \label{eq:example}\\
\nonumber 
\end{align}
Then we can readily check that 
the state 
$\Upsilon_{\bar{D}\bar{A}\bar{B}\widetilde{D}\widetilde{A}\widetilde{B}}
=\ket{\Upsilon}\bra{\Upsilon}$ 
is a QCR state, but not a genuine secret sharing state in Reference~\cite{CL21}.
Furthermore, when $N=1$, the QCR state $\Upsilon_{\bar{D}\bar{A}\widetilde{D}\widetilde{A}}$ 
in Definition~\ref{Def:QCR} can be written as 
\begin{equation}
\frac{1}{d}
\sum_{i, j \in \mathbb{Z}_{d}}
\ket{i, -i}_{\bar{D}\bar{A}}\bra{j, -j}
\otimes
\left(U^{i}_{\widetilde{D}\widetilde{A}}\right)
\sigma_{\widetilde{D}\widetilde{A}}
\left(U^{j}_{\widetilde{D}\widetilde{A}}\right)^{\dagger},
\label{eq:private}
\end{equation}
which is essentially equivalent to a private state, 
and all genuine secret sharing states in Reference~\cite{CL21} 
are QCR states.
Hence, the QCR state can be regarded
as a generalization of the private states and the genuine secret sharing states
with respect to quantum cryptography.


\begin{theorem}
\label{thm1}
Suppose that a dealer and $N$ players share a quantum state $\rho_{D\mathbf{A}}$. 
The dealer and players can obtain information satisfying the above cryptographic conditions~(\ref{condition1}) and~(\ref{condition2}) 
after they measure their information parts in the computational basis
if and only if the state $\rho_{D\mathbf{A}}$ is a QCR state.
\end{theorem}


\begin{theorem}
\label{thm2}
Assume that a dealer $D$ and $N$ players $\mathbf{A}$ share an $(N+1)$-party QCR state.
For any bipartite split $\{\mathbf{P}_{1},\mathbf{P}_{2}\}$
of the players $\mathbf{A}=\mathbf{P}_{1}\mathbf{P}_{2}$ with $\left|\mathbf{P}_{1}\right|=M \ge 1$,
if players $\mathbf{P}_{2}$ measure their information parts
and correctly announce the measurement outcomes,
then $D\mathbf{P}_{1}$ can share an $(M+1)$-party QCR state 
after the dealer applies a proper unitary operation on the dealer's part.
\end{theorem}

\begin{figure}
\centering
\includegraphics[width=.7\linewidth]{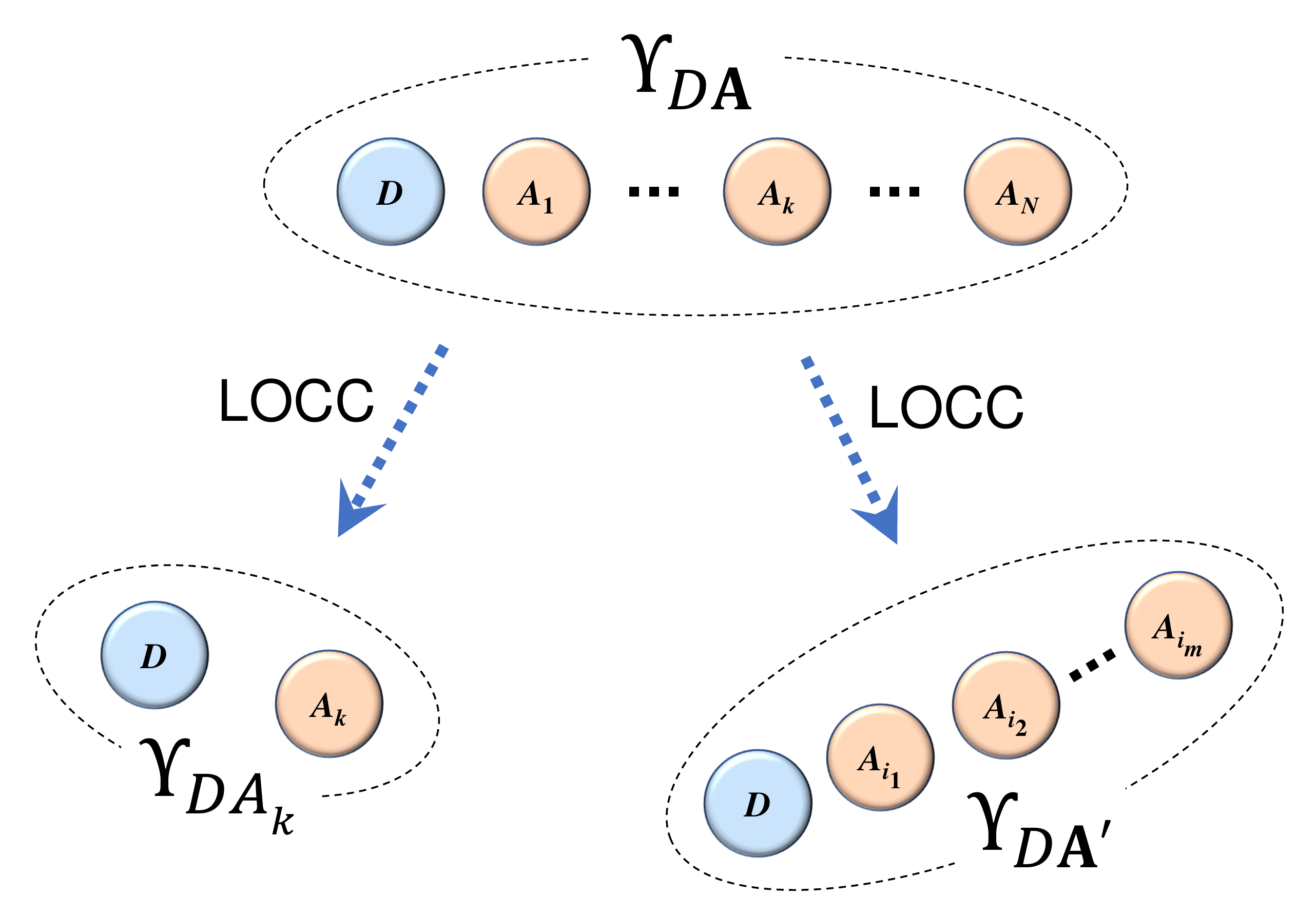}
\caption{
As in Theorem~\ref{thm2},
from the  $\Upsilon_{D\mathbf{A}}$, 
a private state $\Upsilon_{DA_k}$ or a QCR state $\Upsilon_{D\mathbf{A}'}$ with smaller number of parties 
can be obtained by LOCC, 
where $\mathbf{A}'=A_{i_1}A_{i_2}\cdots A_{i_m}$.
}
\label{fig:QCR}
\end{figure}

Theorem~\ref{thm2} tells us that
from a given QCR state, 
a QCR state on any smaller number of players and the dealer 
as well as a private state between any player and the dealer 
can be shared
by LOCC,
as seen in  the Figure~\ref{fig:QCR}.
In other words, Theorem~\ref{thm2} implies that any QCR state satisfies the cryptographic condition~(\ref{condition3}).

\begin{theorem}
\label{thm3}
Assume that there are two QCR states $\Upsilon_{D_A\mathbf{A}}$ and $\Upsilon_{D_B\mathbf{B}}$, 
where both $D_A$ and $D_B$ are the dealer's parties, 
and $\mathbf{A}=A_1\cdots A_N$ and $\mathbf{B}=B_1\cdots B_M$ are 
two different sets of players. 
Then the dealer and all players share a QCR state $\Upsilon_{D\mathbf{AB}}$ via the dealer's proper local operations,
where $D=D_AD_B$.
\end{theorem}

\begin{figure}
\centering
\includegraphics[width=.7\linewidth]{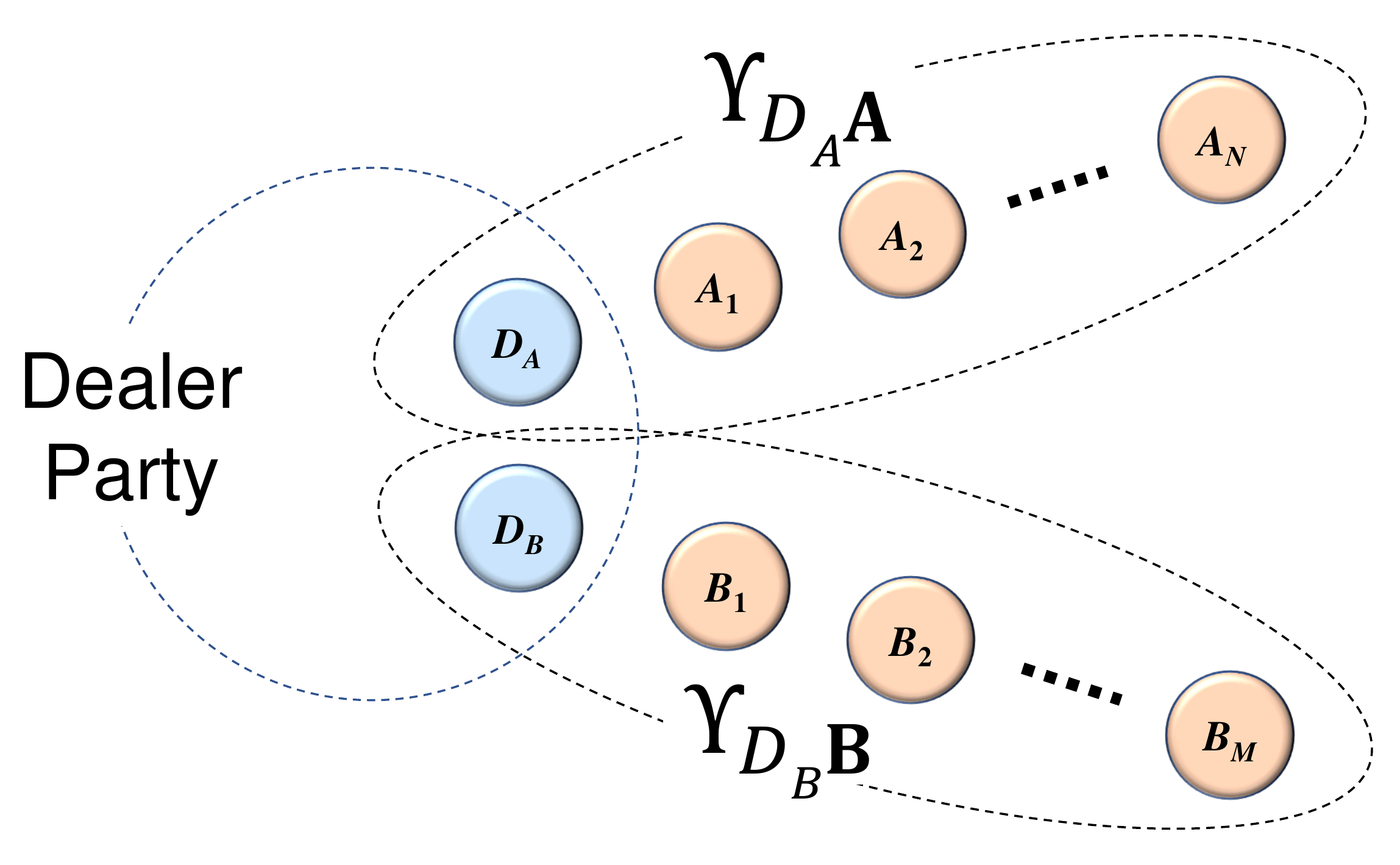}
\caption{
Constructing a QCR state with larger number of parties from two QCR states $\Upsilon_{D_A\mathbf{A}}$ and $\Upsilon_{D_B\mathbf{B}}$ in Theorem~\ref{thm3}.
}
\label{fig:larger_QCR}
\end{figure}

By Theorem~\ref{thm3}, we can see that a larger QCR state can be obtained from two different QCR states with the same dealer party as seen in the Figure~\ref{fig:larger_QCR}. 
Furthermore, we note that the private state is considered as a QCR state with one dealer and one player.
Hence, by mathematical induction, we have the following corollary.

\begin{corollary}
\label{coro4}
Suppose that each of $N$ players shares a private state with one dealer.
Then they can have an $(N+1)$-party QCR state by applying the dealer's proper local operations.
\end{corollary}

\section{QCR distillable rate and bound entangled states}
\label{Sec:QCR_rate}

Before defining the QCR distillable rate,
we look at the distillable entanglement and the distillable key rate.
Let $\Phi^{m}_{AB}$ and $\gamma^{m}_{ABA'B'}$ be denoted by the maximally entangled state with $m=\log\dim(A)=\log\dim(B)$
and the private state with $m=\log\dim(A)=\log\dim(B)$, respectively, where $\dim(\cdot)$ is the dimension of the system.
The distillable entanglement $E_\mathrm{D}$ is defined as the rate at which maximally entangled states
can be distilled under LOCC~\cite{BBPSSW96}, that is, 
\begin{equation}
E_\mathrm{D}\left(\rho_{AB}\right)=\lim_{\delta \rightarrow 0}\lim_{n \rightarrow \infty}
\sup_{\Lambda_{{A}:{B}}}
\left\{E:\|\Lambda\left(\rho_{AB}^{\otimes n}\right)-\Phi_{AB}^{nE}\|_{1}\le\delta\right\},
\label{eq:E_D}
\end{equation}
where $\Lambda_{{A}:{B}}$ is an LOCC protocol between Alice and Bob.
Similarly, the distillable key rate $K_\mathrm{D}$ is defined as the rate at which private states can be distilled under LOCC\cite{HO05,H09}, 
that is, 
\begin{equation}
K_{\mathrm{D}}\left(\rho_{AB}\right)=\lim_{\delta \rightarrow 0}\lim_{n \rightarrow \infty}
\sup_{\Lambda_{{A}:{B}}}
\left\{K:\big\|\Lambda\left(\rho_{AB}^{\otimes n}\right)-\gamma_{AB}^{nK}\big\|_{1}\le\delta\right\}.
\label{eq:K_D}
\end{equation}

Since we can define the QCR distillable rate for any state in a similar way to the above definitions, 
from the definition, we can know 
how many copies of the given state are required to asymptotically distill a QCR state through LOCC.
The QCR distillable rate $CR_\mathrm{D}$ of a given multipartite quantum state $\rho_{D\mathbf{A}}$ is defined as 
\begin{equation}
CR_\mathrm{D}\left(\rho_{D\mathbf{A}}\right)=\lim_{\delta \rightarrow 0}\lim_{n \rightarrow \infty}\sup_{\Lambda}
\left\{K:\big\|\Lambda\left(\rho_{D\mathbf{A}}^{\otimes n}\right)-\Upsilon_{D\mathbf{A}}^{nK}\big\|_{1}\le\delta\right\},
\label{eq:CR_D}
\end{equation}
where $\Lambda$ is the dealer's and all players' LOCC operation, 
and $\Upsilon_{D\mathbf{A}}^{m}$ denotes a QCR state 
whose information part $\bar{D}\bar{A}_{1}\bar{A}_{2}\cdots \bar{A}_{N}$ satisfies $m=\log\dim(\bar{D})=\log\dim(\bar{A}_{i})$ for all $i$.

Let us now investigate the connection between the distillable key rate and the QCR distillable rate.
It follows from Theorem~\ref{thm2} that 
\begin{equation}
CR_{\mathrm{D}}\left(\rho_{D\mathbf{A}}\right) \le 
K^{D\mathbf{P}_{1}:\mathbf{P}_{2}}_{\mathrm{D}}\left(\rho_{D\mathbf{A}}\right) 
\label{eq:CR_D_K_D}
\end{equation}
for any bipartite split $\{\mathbf{P}_{1}, \mathbf{P}_{2}\}$ of the players  
$\mathbf{A}=\mathbf{P}_1\mathbf{P}_2$.
In addition, by Theorem~\ref{thm3}, we have the following theorem. 

\begin{theorem}
\label{thm5}
Let $\mathbf{A}=A_1\cdots A_N$ and $\mathbf{B}=B_1\cdots B_M$ be 
two different sets of players, 
and let $D_A$ and $D_B$ be the dealer's parties.  
For given two states $\rho_{D_A\mathbf{A}}$ and ${\rho}_{D_B\mathbf{B}}$,
\begin{equation}
CR_{\mathrm{D}}\left(\rho_{D_A\mathbf{A}}\otimes\rho_{D_B\mathbf{B}} \right) 
\ge \mathrm{min}\{CR_{\mathrm{D}}\left(\rho_{D_A\mathbf{A}}\right),
CR_{\mathrm{D}}\left(\rho_{D_B\mathbf{B}}\right)\}. \label{eq:thm5}
\end{equation}

\end{theorem}

Hence, the following corollary clearly comes from Theorem~\ref{thm5} and Corollary~\ref{coro4}.

\begin{corollary}
\label{coro6}
For each $i=1, 2, \ldots, N$, let $\rho_{D_iA_i}$ be the quantum state shared by the dealer $D_i$ and the $i$-th player $A_i$.
Then the following inequality holds.
\begin{equation}
CR_{\mathrm{D}}\left(\bigotimes^N_{i=1} \rho_{D_iA_i}\right)
\ge \mathrm{min}_{i}\left\{K_{\mathrm{D}}(\rho_{D_iA_i})\right\}.
\label{eq:cor6}
\end{equation}
\end{corollary}

Corollary~\ref{coro6} implies that
if each $\rho_{D_iA_i}$ has a positive distillable key rate,
then $\bigotimes^N_{i=1} \rho_{D_iA_i}$ has a positive QCR distillable rate.
We note that 
if each $\rho_{D_iA_i}$ is a bipartite state with positive partial transposition (PPT),
then $\bigotimes^N_{i=1} \rho_{D_iA_i}$ is also an $(N+1)$-partite state with PPT, since it is a PPT state with respect to any bipartite split of $D\mathbf{A}$ with one dealer $D=D_1D_2\cdots D_N$ and $N$ players $\mathbf{A}=A_1A_2\cdots A_N$.
Hence, we can readily construct multipartite PPT bound entangled states 
with positive QCR distillable rate
from bipartite PPT bound entangled states with positive distillable key rate,
which are presented in References~\cite{CL07,H09}. 
Therefore, we can finally present our theorem showing the existence of such states as follows.
\begin{theorem}
\label{coro7}
For any natural number $N\ge 2$,
there exists an $(N+1)$-partite bound entangled state $\rho_{D\mathbf{A}}$ with $CR_\mathrm{D}\left(\rho_{D\mathbf{A}}\right)>0$. 
\end{theorem}

\section{Discussion}

We have defined the QCR state with a dealer party, 
and have shown that a given multipartite quantum state is a QCR state 
if and only if the two cryptographic conditions on the state hold. 
We have also defined the QCR distillable rate of a given multipartite state, 
and have presented several important properties on the QCR distillable rate. 
In the sequel, we have presented how to construct a QCR distillable state with larger number of parties from several QCR distillable states. 
Moreover, we have proved that there exist multipartite bound entangled states which are QCR distillable.
This result implies that there exists a multipartite quantum state on which 
a dealer and players can perform one of several kinds of quantum cryptographic protocols to some extent, 
and from which they cannot distill any bipartite nor multipartite entanglement by LOCC.
Hence, we can conclude that any bipartite or multipartite distillable entanglement is not necessarily required for quantum cryptography.  

The QCR states that we have dealt with in this paper 
have one specific dealer party. 
Thus several kinds of perfectly secure classical communication feasible on the quantum state 
can be performed between the dealer party and any number of players. 
Therefore, the QCR state can be considered as a resource unit in quantum cryptographic theory, 
and hence we could construct the quantum cryptographic network consisting of the QCR states instead of the bipartite maximally entangled states or the private states.

\acknowledgements{
This research was supported by the National Research Foundation of Korea (NRF) grant funded
by the Ministry of Science and ICT (MSIT) (Grants No. NRF-2019R1A2C1006337 and No. NRF-2020M3E4A1079678).
S.L. acknowledges support from the MSIT, Korea, 
under the Information Technology Research Center support program (Grant No. IITP-2021-2018-0-01402) 
supervised by the Institute for Information and Communications Technology Planning and Evaluation, 
and the Quantum Information Science and Technologies program of the NRF 
funded by the MSIT (Grant No. 2020M3H3A1105796).}

\appendix

\section{Proof of Theorem~\ref{thm1}}
This proof is almost the same as that of the theorem related to the genuine secret sharing state in Reference~\cite{CL21}. 
The details are as follows. 

Let us consider the state
\begin{equation}
\ket{\Gamma_\rho}_{\bar{D}\mathbf{\bar{A}}\widetilde{D}\mathbf{\widetilde{A}}E}
=\sum_{I \in \mathbb{Z}_{d}^{N+1}}
\sqrt{p_{I}}\ket{I}_{\bar{D}\mathbf{\bar{A}}}\ket{\psi_{I}}_{\widetilde{D}\mathbf{\widetilde{A}}E},
\label{thm1_purif}
\end{equation}
which is a purification of $\rho_{D\mathbf{A}}$.
Assume that the dealer and players can have cryptographic information that satisfies the cryptographic condition~(\ref{condition1})
by measuring the information part of $\rho_{D\mathbf{A}}$.
Then we have $p_{I}=1/d^{N}$ for $I \in \mathfrak{S}_{N+1}^{0}$
and $p_{I}=0$ for $I \notin \mathfrak{S}_{N+1}^{0}$.

Regarding the condition~(\ref{condition2}),  
we first take account of the worst case that 
all players except one player, say $A_{k}$, are dishonest.
Then the subsystem $\mathbf{\bar{A}}'=\bar{A}_1\cdots\bar{A}_{k-1}\bar{A}_{k+1}\cdots \bar{A}_N$ 
is the information part of the dishonest players. 

Let $i$ be the dealer's measurement outcome.
Then the eavesdropper and dishonest players' state after measurement becomes
\begin{equation}
\gamma_{\mathbf{\bar{A}'}\widetilde{D}\mathbf{\widetilde{A}}E}^{(i)}
=\frac{1}{d^{N-1}}\sum_{i_k\in \mathbb{Z}_{d}}
\sum_{\xi,\xi' \in \mathfrak{S}_{N-1}^{-i-i_k}}
\ket{\xi}_{\mathbf{\bar{A}'}}\bra{\xi'}
\otimes
{\rm{tr}}_{\widetilde{D}\widetilde{A}_{k}}
\ket{\psi_{i,i_k,\xi}}_{\widetilde{D}\mathbf{\widetilde{A}}E}\bra{\psi_{i,i_k,\xi'}}
\label{eq:mixed1}
\end{equation}
if reordering the systems.
From the cryptographic condition~(\ref{condition2}), 
we have $\gamma_{\mathbf{\bar{A}'}\widetilde{D}\mathbf{\widetilde{A}}E}^{(i)}
=\gamma_{\mathbf{\bar{A}'}\widetilde{D}\mathbf{\widetilde{A}}E}^{(i')}$
for any $i,i' \in \mathbb{Z}_{d}$.
It follows from the Hughston-Jozsa-Wootters theorem~\cite{HJW}
that for $i,i_k, i',i_k' \in \mathbb{Z}_{d}$ with $i+i_k=i'+i_k' \pmod{d}$, 
there is a unitary operator $U_{\widetilde{D}\widetilde{A}_{k}}^{i,i_k \to i',i_k'}$
on the system $\widetilde{D}\widetilde{A}_{k}$ such that
\begin{equation}
\label{HJW relation}
U_{\widetilde{D}\widetilde{A}_{k}}^{i,i_k \to i',i_k'}
\ket{\psi_{i,i_k,\xi}}_{\widetilde{D}\mathbf{\widetilde{A}}E}
=\ket{\psi_{i',i_k',\xi}}_{\widetilde{D}\mathbf{\widetilde{A}}E}
\end{equation}
for all $\xi \in \mathfrak{S}_{N-1}^{-i-i_k}$.

Let  $\{\mathbf{P}_1, \mathbf{P}_2\}$ be an arbitrary bipartite split  of the players 
with $\mathbf{P}_1$ consisting of at least one player and $\mathbf{A}= \mathbf{P}_1\mathbf{P}_2$.
Without loss of generality,
we may assume that $\mathbf{P}_1=A_{1}\cdots{A}_{M}$ and
$\mathbf{P}_{2}={A}_{M+1}\cdots{A}_{N}$.
Then by Eq.~(\ref{HJW relation}),
it can be shown that
if $iI_{1}I_{2}=ii_{1}\cdots i_{M} i_{M+1} \cdots i_{N} \in \mathfrak{S}_{N+1}^{0}$,
then
\begin{equation}
\label{thm1_relation}
\ket{\psi_{00\cdots 0}}_{\mathbf{A'}E}
=U_{\widetilde{D}\widetilde{A}_{N}}^{j_{N-1},i_{N}}
\cdots
U_{\widetilde{D}\widetilde{A}_{2}}^{j_{1},i_{2}}
U_{ \widetilde{D}\widetilde{A}_{1}}^{i,i_{1}}
\ket{\psi_{ii_{1}i_{2}\cdots i_{N}}}_{\mathbf{A'}E},
\end{equation}
where $U_{\widetilde{D}\widetilde{A}_{k}}^{i,j}=U_{\widetilde{D}\widetilde{A}_{k}}^{i,j \rightarrow i+j,0}$
and $j_{t}\equiv i+i_{1}+\cdots +i_{t} \pmod{d}$.
Let 
${\rm{tr}}_{\widetilde{D}\mathbf{\widetilde{A}}}\left( \ket{\psi_{00\cdots 0}}\bra{\psi_{00\cdots 0}}\right)
=\sum_{x}\lambda_{x}\ket{\eta_{x}}_{E}\bra{\eta_{x}}$
be its spectral decomposition.
Then we have
\begin{equation}
\ket{\psi_{iI_{1}I_{2}}}_{\widetilde{D}\mathbf{\widetilde{A}}E}
=\sum_{x}\sqrt{\lambda_{x}}
U_{\widetilde{D}\mathbf{\widetilde{P}}_{1}}^{iI_{1}}V_{\widetilde{D}\mathbf{\widetilde{A}}}^{I_{2}}
\ket{\phi_{x}}_{\widetilde{D}\mathbf{\widetilde{A}}}\ket{\eta_{x}}_{E}
\label{eq:final_psi}
\end{equation}
for some  unitary operators $U_{\widetilde{D}\mathbf{\widetilde{P}}_{1}}^{iI_{1}}$, $V_{\widetilde{D}\mathbf{\widetilde{A}}}^{I_{2}}$
and orthonormal set $\{\ket{\phi_{x}}\}$ for the system $\widetilde{D}\mathbf{\widetilde{A}}$.
Therefore, $\rho_{D\mathbf{A}}$ is of the form in Eq.~(\ref{QCR_state}).

Conversely, assume that $\rho_{D\mathbf{A}}$ has the form in Eq.~(\ref{QCR_state}).
Then it can be readily shown that players have cryptographic information that obeys the cryptographic condition (\ref{condition1})
after measuring their information parts in the computational basis.

We now show that players' cryptographic information satisfies the condition (\ref{condition2}).
Suppose that 
$\{\mathbf{P}_1, \mathbf{P}_2\}$ is a bipartite split  of the players $\mathbf{A}= \mathbf{P}_1\mathbf{P}_2$
with $\mathbf{P}_1$ consisting of at least one player 
and $\mathbf{P}_{2}$ representing $K$ dishonest players.
Let $\sigma_{\widetilde{D}\mathbf{\widetilde{A}}}=\sum_{x}\kappa_{x}\ket{\mu_{x}}_{\widetilde{D}\mathbf{\widetilde{A}}}\bra{\mu_{x}}$
be a spectral decomposition of $\sigma_{\widetilde{D}\mathbf{\widetilde{A}}}$,
and let
\begin{equation}
\ket{\varphi_{iI_{1}I_{2}}}_{\widetilde{D}\mathbf{\widetilde{A}}E}
=\sum_{x}\sqrt{\kappa_{x}}
U_{\widetilde{D}\mathbf{\widetilde{P}}_{1}}^{iI_{1}}V_{\widetilde{D}\mathbf{\widetilde{A}}}^{I_{2}}
\ket{\mu_{x}}_{\widetilde{D}\mathbf{\widetilde{A}}}\ket{\nu_{x}}_{E},
\label{eq:varphi}
\end{equation}
where $\left\{\ket{\nu_{x}}\right\}$ forms an orthonormal set for the eavesdropper's system $E$.
Then the state 
\begin{equation}
\ket{\Upsilon}_{D\mathbf{P_{1}}\mathbf{P_{2}}E}
=\frac{1}{\sqrt{d^{N}}}\sum_{iI_{1}I_{2} \in \mathfrak{S}_{N+1}^{0}}
\ket{iI_{1}}_{\bar{D}\mathbf{\bar{P}_{1}}}\ket{I_{2}}_{\mathbf{\bar{P}_{2}}}\ket{\varphi_{iI_{1}I_{2}}}_{\widetilde{D}\mathbf{\widetilde{A}}E}
\label{eq:purification_of_rho}
\end{equation}
is a purification of $\rho_{D\mathbf{A}}$.
If the dealer has the measurement outcome $i$
after measuring the dealer's information part in the computational basis,
then the quantum state of dishonest players and eavesdropper after the measurement becomes
\begin{equation}
\Upsilon_{\mathbf{P}_{2}E}^{(i)}
=\frac{1}{d^{K}}\sum_{\alpha \in \mathbb{Z}_{d}}\sum_{I_{2},J_{2} \in \mathfrak{S}_{K}^{-\alpha}}
\ket{I_{2}}_{\mathbf{\bar{P}}_{2}}\bra{J_{2}} \otimes
\mathrm{tr}_{\widetilde{D}\mathbf{\widetilde{P}}_{1}}\ket{\varphi_{I_{2}}}_{\widetilde{D}\mathbf{\widetilde{P}}E}\bra{\varphi_{J_{2}}},
\label{eq:condition2state}
\end{equation}
where $\ket{\varphi_{I_{2}}}_{\widetilde{D}\mathbf{\widetilde{P}}E}
=\sum_{x}\sqrt{\kappa_{x}}V_{\widetilde{D}\mathbf{\widetilde{P}}}^{I_{2}}
\ket{\xi_{x}}_{\widetilde{D}\mathbf{\widetilde{P}}}\ket{e_{x}}_{E}.$
Since $\Upsilon_{\mathbf{P}_{2}E}^{(i)}
=\Upsilon_{\mathbf{P}_{2}E}^{(j)}$
for any $i, j \in \mathbb{Z}_{d}$, 
dishonest players and eavesdropper
cannot get any information about the dealer's cryptographic information.

\section{Proof of Theorem~\ref{thm2}}
The proof of Theorem~\ref{thm2} is also similar to that of the theorem associated with the genuine secret sharing states in Reference~\cite{CL21}.
However, we here present its simple proof compared to that in Reference~\cite{CL21} as follows. 

Without loss of generality,
we may assume that $\mathbf{P}_{1}={A}_{1}\cdots{A}_{M}$ and
$\mathbf{P}_{2}={A}_{M+1}\cdots{A}_{N}$.
Let $\Upsilon_{D\mathbf{A}}$ be an $(N+1)$-party QCR state
shared by a dealer and $N$ players.
Since $\Upsilon_{D\mathbf{A}}$ has the form in Eq.~(\ref{QCR_state}),
if let  $I_{2} \in \mathfrak{S}_{N-M}^{\beta}$ be the measurement outcomes for some $\beta$
when players $\mathbf{P}_{2}$ measure their information parts in the computational basis, 
then the resulting state of the dealer $D$ and the players $\mathbf{P}_{1}$ after the measurement becomes
\begin{equation}
\Upsilon_{D\mathbf{P}_{1}}^{(I_{2})}
=\frac{1}{d^{M}}\sum_{iI_{1},jJ_{1} \in \mathfrak{S}_{M+1}^{-\beta}}
\ket{iI_{1}}_{\bar{D}\mathbf{\bar{P}}_{1}}\bra{jJ_{1}} \otimes
U_{\widetilde{D}\mathbf{\widetilde{P}}_{1}}^{iI_{1}}
\tilde{\sigma}_{\widetilde{D}\mathbf{\widetilde{P}}_{1}}
\left(U_{\widetilde{D}\mathbf{\widetilde{P}}_{1}}^{jJ_{1}}\right)^{\dagger},
\label{eq:proof2}
\end{equation}
where $\tilde{\sigma}_{\widetilde{D}\mathbf{\widetilde{P}}_1}=
\mathrm{tr}_{\mathbf{\widetilde{P}}_{2}}\tilde{V}_{\widetilde{D}\mathbf{\widetilde{A}}}^{I_{2}}
\sigma_{\widetilde{D}\mathbf{\widetilde{A}}}
\left(\tilde{V}_{\widetilde{D}\mathbf{\widetilde{A}}}^{I_{2}}\right)^{\dagger}$.

We note that unitary operators on the shield part of the state $\Upsilon_{D\mathbf{A}}$
can be expressed as in Eq.~(\ref{thm1_relation}), 
and it can be easily shown that
$W_{\bar{D}}\Upsilon_{D\mathbf{P}_{1}}^{(I_{2})}W_{\bar{D}}^{\dagger}$
is an $(M+1)$-party QCR state,
where $W=\sum_{i=0}^{d-1}\ket{i+\beta}\bra{i}$.
Therefore, if the players $\mathbf{P}_{2}$ announces the value $\beta$,
then the dealer $D$ and the players $\mathbf{P}_{1}$  can share the $(M+1)$-party QCR state
after applying the unitary operator $W$ on the dealer's information part.

\section{Proof of Theorem~\ref{thm3}}
Let
\begin{equation}
\ket{\Upsilon}_{D_A\mathbf{A}E_{A}}=
\frac{1}{\sqrt{d^{N}}}
\sum_{iI \in \mathfrak{S}_{N+1}^{0}}
\ket{iI}_{\bar{D}_A\bar{\mathbf{A}}}
\otimes
\ket{\psi_{iI}}_{\widetilde{D}_A\mathbf{\widetilde{A}}E_{A}}
\label{eq:DA}
\end{equation}
be a purification of the QCR state $\Upsilon_{D_A\mathbf{A}}$,
and let
\begin{equation}
\ket{\Upsilon}_{D_B\mathbf{B}E_{B}}=
\frac{1}{\sqrt{d^{M}}}
\sum_{jJ \in \mathfrak{S}_{M+1}^{0}}
\ket{jJ}_{\bar{D}_B\bar{\mathbf{B}}}
\otimes
\ket{\phi_{jJ}}_{\widetilde{D}_B\mathbf{\widetilde{B}}E_{B}}
\label{eq:DB}
\end{equation}
be a purification of the QCR state $\Upsilon_{D_B\mathbf{B}}$.
For $I=i_1i_2\cdots i_L \in \mathbb{Z}_d^{L}$,   
let $|I|$ be defined as $|I|=i_{1}+\cdots+i_{L}$.
Then the states $\ket{\Upsilon}_{D_A\mathbf{A}E_{A}}$ 
and $\ket{\Upsilon}_{D_B\mathbf{B}E_{B}}$ in Eqs.~(\ref{eq:DA}) and (\ref{eq:DB}) can be rewritten as
\begin{equation}
\ket{\Upsilon}_{D_A\mathbf{A}E_{A}}=
\frac{1}{\sqrt{d^{N}}}
\sum_{I \in \mathbb{Z}_{d}^{N}}
\ket{-|I|,I}_{\bar{D}_A\bar{\mathbf{A}}}
\otimes
\ket{\psi_{-|I|,I}}_{\widetilde{D}_A\mathbf{\widetilde{A}}E_{A}}
\label{eq:DA2}
\end{equation}
and
\begin{equation}
\ket{\Upsilon}_{D_B\mathbf{B}E_{B}}=
\frac{1}{\sqrt{d^{M}}}
\sum_{J \in \mathbb{Z}_{d}^{M}}
\ket{-|J|,J}_{\bar{D}_B\bar{\mathbf{B}}}
\otimes
\ket{\phi_{-|J|,J}}_{\widetilde{D}_B\mathbf{\widetilde{B}}E_{B}}
\label{eq:DB2}
\end{equation}
respectively. 

Let $cX$ be the unitary operator defined as 
\begin{equation}
cX=\sum_{j,k=0}^{d-1}\ket{j+k,k}\bra{j,k}.
\label{eq:cX}
\end{equation}
If the dealer applies the unitary operator $cX_{\bar{D}_{A}\bar{D}_{B}}$
on the system $\bar{D}_{A}\bar{D}_{B}$ 
in the state $\ket{\Upsilon}_{D_A\mathbf{A}E_{A}} \otimes \ket{\Upsilon}_{D_B\mathbf{B}E_{B}}$,
then after properly rearranging the order of the systems, the state becomes
\begin{align}
\frac{1}{\sqrt{d^{N+M}}}
\sum_{I \in \mathbb{Z}_{d}^{N}}
\sum_{J \in \mathbb{Z}_{d}^{M}}&
\ket{-|I|-|J|,I,J}_{\bar{D}_{A}{\mathbf{\bar{A}}}{\mathbf{\bar{B}}}} 
\otimes
\ket{-|J|}_{\bar{D}_{B}} 
\otimes \ket{\psi_{-|I|,I}}_{\widetilde{D}_A\mathbf{\widetilde{A}}E_{A}} 
\otimes \ket{\phi_{-|J|,J}}_{\widetilde{D}_B\mathbf{\widetilde{B}}E_{B}}.
\label{eq:DADB} \\
\nonumber
\end{align}
We remark that if the dealer and all players measure their information part $\bar{D}_{A}{\mathbf{\bar{A}}}{\mathbf{\bar{B}}}$ in the computational basis,
then they have cryptographic information that satisfies the cryptographic condition~(\ref{condition1}).
In order to show that the cryptographic information obeys the cryptographic condition~(\ref{condition2}),
we consider the worst case as in the proof of Theorem~\ref{thm1}.

Let us assume that the dealer measures the information part $\bar{D}_{A}$, and 
let $i$ be the dealer's measurement outcome.
By tracing out the system $\bar{D}_{A}\bar{D}_{B}\widetilde{D}_A\widetilde{D}_B$ of the resulting state,
we have
\begin{align}
\frac{1}{d^{N+M-1}}
\sum_{\alpha \in \mathbb{Z}_{d}}
\sum_{I, I' \in \mathfrak{S}_{N}^{\alpha}}\sum_{J, J' \in \mathfrak{S}_{M}^{-\alpha-i}}&
\ket{I,J}_{\bar{\mathbf{A}}\bar{\mathbf{B}}} \bra{I',J'}
\otimes \mathrm{tr}_{\widetilde{D}_A}\ket{\psi_{-\alpha,I}}_{\widetilde{D}_A\mathbf{\widetilde{A}}E_{A}}  \bra{\psi_{-\alpha,I'}}
\otimes \mathrm{tr}_{\widetilde{D}_B}\ket{\phi_{\alpha+i,J}}_{\widetilde{D}_B\mathbf{\widetilde{B}}E_{B}} \bra{\phi_{\alpha+i,J'}}.
\label{eq:DADB3}
\\
\nonumber
\end{align}
Let us now consider the situation 
where all players except the dealer and one player
are dishonest as the worst case.
Without loss of generality, 
we may assume that the honest player is $A_1$, by symmetry. 
When $N \ge 2$, after tracing out the system $A_{1}$,
the dishonest players and eavesdropper's state becomes
\begin{align}
\frac{1}{d^{N+M-1}}
\sum_{\alpha, \beta \in \mathbb{Z}_{d}}
\sum_{\hat{I}, \hat{I}' \in \mathfrak{S}_{N-1}^{\alpha-\beta}}\sum_{J, J' \in \mathfrak{S}_{M}^{-\alpha-i}}&
\ket{\hat{I}}_{\hat{\mathbf{A}}} \bra{\hat{I}'}\otimes\ket{J}_{\bar{\mathbf{B}}} \bra{J'} 
\nonumber\\
&\otimes \mathrm{tr}_{\widetilde{D}_{A}\widetilde{A}_{1}}
\ket{\psi_{-\alpha,\beta,\hat{I}}}_{\widetilde{D}_A\mathbf{\widetilde{A}}E_{A}}  \bra{\psi_{-\alpha,\beta,\hat{I}'}}
\otimes \mathrm{tr}_{\widetilde{D}_B}\ket{\phi_{\alpha+i,J}}_{\widetilde{D}_B\mathbf{\widetilde{B}}E_{B}} \bra{\phi_{\alpha+i,J'}},
\label{eq:DADB4}
\\
\nonumber
\end{align}
where $\hat{I}=i_2\cdots i_L\in\mathbb{Z}_d^{L-1}$ for $I=i_1i_2\cdots i_L\in\mathbb{Z}_d^L$
and 
$\hat{\mathbf{A}}=\bar{A}_2\cdots \bar{A}_L$ for $\bar{\mathbf{A}}=\bar{A}_1\bar{A}_2\cdots \bar{A}_L$.
Since $\Upsilon_{D_A\mathbf{A}}$ is a QCR state,
\begin{equation}
\mathrm{tr}_{\widetilde{D}_{A}\widetilde{A}_{1}}
\ket{\psi_{-\alpha,\beta,\hat{I}}}_{\widetilde{D}_A\mathbf{\widetilde{A}}E_{A}}  \bra{\psi_{-\alpha,\beta,\hat{I}'}}
=\mathrm{tr}_{\widetilde{D}_{A}\widetilde{A}_{1}}
\ket{\psi_{0,\beta-\alpha,\hat{I}}}_{\widetilde{D}_A\mathbf{\widetilde{A}}E_{A}}  \bra{\psi_{0,\beta-\alpha,\hat{I}'}}
\label{eq:equal}
\end{equation}
for any $\alpha,\beta \in \mathbb{Z}_{d}$ and $\hat{I}, \hat{I}' \in \mathfrak{S}_{N-1}^{\alpha-\beta}$.
Hence, the state in Eq.~(\ref{eq:DADB4}) can be rewritten as
\begin{align}
\frac{1}{d^{N+M-1}}
\sum_{s, t \in \mathbb{Z}_{d}}
\sum_{\hat{I}, \hat{I}' \in \mathfrak{S}_{N-1}^{s}}\sum_{J, J' \in \mathfrak{S}_{M}^{t}}&
\ket{\hat{I}}_{\hat{\mathbf{A}}} \bra{\hat{I}'}\otimes\ket{J}_{\bar{\mathbf{B}}} \bra{J'} 
\nonumber\\
&\otimes \mathrm{tr}_{\widetilde{D}_{A}\widetilde{A}_{1}}
\ket{\psi_{0,-s,\hat{I}}}_{\widetilde{D}_A\mathbf{\widetilde{A}}E_{A}}  \bra{\psi_{0,-s,\hat{I}'}}
\otimes \mathrm{tr}_{\widetilde{D}_B}\ket{\phi_{-t,J}}_{\widetilde{D}_B\mathbf{\widetilde{B}}E_{B}} \bra{\phi_{-t,J'}}.
\label{eq:DADB5}
\\
\nonumber
\end{align}
We can here see that the state in Eq.~(\ref{eq:DADB5}) is independent on the dealer's measurement outcome $i$.
In other words, the dealer's cryptographic information is perfectly secure against the dishonest players and any exterior eavesdropper,
which implies that the dealer's and all players' cryptographic information satisfies the cryptographic condition (\ref{condition2}).

Now assume that $N=1$, that is, $\mathbf{A}=A_1$. 
Then the state of the dishonest players $\mathbf{B}$ and eavesdropper $E_AE_B$ after the dealer's measurement is
\begin{align}
\frac{1}{d^{M}}
\sum_{\alpha \in \mathbb{Z}_{d}}
\sum_{J, J' \in \mathfrak{S}_{M}^{-\alpha-i}}&
\ket{J}_{\bar{\mathbf{B}}} \bra{J'}
\otimes \mathrm{tr}_{\widetilde{D}_{A}\widetilde{\mathbf{A}}}\ket{\psi_{-\alpha,\alpha}}_{\widetilde{D}_A\mathbf{\widetilde{A}}E_{A}}  \bra{\psi_{-\alpha,\alpha}}
\otimes \mathrm{tr}_{\widetilde{D}_B}\ket{\phi_{\alpha+i,J}}_{\widetilde{D}_B\mathbf{\widetilde{B}}E_{B}} \bra{\phi_{\alpha+i,J'}},
\label{eq:DDB}\\
\nonumber
\end{align}
where the dealer's measurement outcome is $i$. 
Since 
\begin{equation}
\mathrm{tr}_{\widetilde{D}_{A}\widetilde{\mathbf{A}}}\ket{\psi_{-\alpha,\alpha}}_{\widetilde{D}_A\mathbf{\widetilde{A}}E_{A}}  \bra{\psi_{-\alpha,\alpha}}
=\mathrm{tr}_{\widetilde{D}_{A}\widetilde{\mathbf{A}}}\ket{\psi_{0,0}}_{\widetilde{D}_A\mathbf{\widetilde{A}}E_{A}}  \bra{\psi_{0,0}}
\label{eq:equal2}
\end{equation}
for all $\alpha \in \mathbb{Z}_{d}$,
the state in Eq.~(\ref{eq:DDB}) does not depend on the measurement outcome $i$,
and hence the cryptographic information is perfectly secure against the dishonest players and any exterior eavesdropper. 

Let $\ket{\Upsilon}_{D\mathbf{AB}E}$ be the pure state in Eq.~(\ref{eq:DADB}),
which is the resulting state after the dealer applies the unitary operator $cX_{\bar{D}_{A}\bar{D}_{B}}$
on the system $\bar{D}_{A}\bar{D}_{B}$ 
in the state $\ket{\Upsilon}_{D_A\mathbf{A}E_{A}} \otimes \ket{\Upsilon}_{D_B\mathbf{B}E_{B}}$,
where $D=D_AD_B$ and $E=E_AE_B$. 
Then, for any cases, the cryptographic information from the state $\ket{\Upsilon}_{D\mathbf{AB}E}$
obeys the cryptographic conditions~(\ref{condition1}) and (\ref{condition2}).
Therefore, the state $cX_{\bar{D}_{A}\bar{D}_{B}}\left(\Upsilon_{D_A\mathbf{A}} \otimes \Upsilon_{D_B\mathbf{B}}\right)cX^\dagger_{\bar{D}_{A}\bar{D}_{B}}$
is an $(N+M+1)$-party QCR state by Theorem~\ref{thm1}, 
since the state is equal to 
$\mathrm{tr}_E\ket{\Upsilon}_{D\mathbf{AB}E}\bra{\Upsilon}$.

\section{Proof of Theorem~\ref{thm5}}
We first note that the set of all LOCC operations $\Lambda_{D\mathbf{AB}}$ on the dealer $D=D_AD_B$ and all players $\mathbf{AB}$
contains LOCC operations of the form $\Lambda_{D_A\mathbf{A}}\otimes{\Lambda}_{D_B\mathbf{B}}$.
Hence, $CR_\mathrm{D}\left(\rho_{D_A\mathbf{A}}\otimes{\rho}_{D_B\mathbf{B}} \right)$ is lower bounded by
\begin{equation}
\lim_{\delta \rightarrow 0}\lim_{n \rightarrow \infty}
\sup_{U_{D}}\sup_{\Lambda_{D_A\mathbf{A}}, {\Lambda}_{D_B\mathbf{B}}}
\left\{K:\Big\|U_{D}\left(\Lambda_{D_A\mathbf{A}}\left(\rho^{\otimes n}_{D_A\mathbf{A}}\right)
\otimes {\Lambda}_{D_B\mathbf{B}}\left({\rho}_{D_B\mathbf{B}}^{\otimes n}\right)\right)U^\dagger_D
-\Upsilon_{D\mathbf{AB}}^{nK}\Big\|_{1}\le\delta\right\},
\label{lowerbound1}
\end{equation}
where $U_{D}$'s are unitary operators acting on the system $D$. 
In addition, as seen in the proof of Theorem~\ref{thm3}, 
there exists 
a unitary operator $\bar{U}_D$ such that 
\begin{equation}
\bar{U}_D\left(\Upsilon^{nK}_{D_A\mathbf{A}}\otimes{\Upsilon}^{nK}_{D_B\mathbf{B}} \right)\bar{U}^\dagger_D = \Upsilon^{nK}_{D\mathbf{AB}}.
\label{eq:U_D}
\end{equation}
Then it follows from Eq.~(\ref{eq:U_D}) that
\begin{equation}
\Big\|\bar{U}_{D}\left(\Lambda_{D_A\mathbf{A}}\left(\rho^{\otimes n}_{D_A\mathbf{A}}\right)
\otimes {\Lambda}_{D_B\mathbf{B}}\left({\rho}_{D_B\mathbf{B}}^{\otimes n}\right)\right)\bar{U}^\dagger_{D}
-\Upsilon_{D\mathbf{AB}}^{nK}\Big\|_{1}
=
\Big\|\Lambda_{D_A\mathbf{A}}\left(\rho^{\otimes n}_{D_A\mathbf{A}}\right)
\otimes {\Lambda}_{D_B\mathbf{B}}\left({\rho}_{D_B\mathbf{B}}^{\otimes n}\right)
-\Upsilon^{nK}_{D_A\mathbf{A}}\otimes{\Upsilon}^{nK}_{D_B\mathbf{B}} \Big\|_{1}.
\label{eq:equal3}   
\end{equation}
 
By the telescoping property of the trace distance~\cite{Wilde0,Wilde},
that is, \begin{equation}
\|\sigma_{1}\otimes\sigma_{2}-\tau_{1}\otimes\tau_{2}\|_{1}
\le \|\sigma_{1}-\tau_{1}\|_{1}+\|\sigma_{2}-\tau_{2}\|_{1},
\end{equation}
we can see that 
the lower bound on $CR_\mathrm{D}\left(\rho_{D_A\mathbf{A}}\otimes{\rho}_{D_B\mathbf{B}} \right)$ in Eq.~(\ref{lowerbound1}) is also lower bounded by
\begin{equation}
\lim_{\delta \rightarrow 0}\lim_{n \rightarrow \infty}
\sup_{\Lambda_{D_A\mathbf{A}}, \Lambda_{D_B\mathbf{B}}}
\left\{K:\Big\|\Lambda_{D_A\mathbf{A}}\left(\rho^{\otimes n}_{D_A\mathbf{A}}\right)
-\Upsilon^{nK}_{D_A\mathbf{A}} \Big\|_{1}
\le\frac{\delta}{2}, 
\Big\|
{\Lambda}_{D_B\mathbf{B}}\left({\rho}_{D_B\mathbf{B}}^{\otimes n}\right)
-{\Upsilon}^{nK}_{D_B\mathbf{B}} \Big\|_{1}
\le\frac{\delta}{2}\right\},
\label{lowerbound2}
\end{equation}
which is greater than or equal to 
both $CR_{\mathrm{D}}\left(\rho_{D_A\mathbf{A}}\right)$ and
$CR_{\mathrm{D}}\left(\rho_{D_B\mathbf{B}}\right)$.
This completes the proof, that is,  
\begin{equation}
CR_{\mathrm{D}}\left(\rho_{D_A\mathbf{A}}\otimes\rho_{D_B\mathbf{B}} \right) 
\ge \mathrm{min}\{CR_{\mathrm{D}}\left(\rho_{D_A\mathbf{A}}\right),
CR_{\mathrm{D}}\left(\rho_{D_B\mathbf{B}}\right)\}.
\label{eq:final}
\end{equation}

\bibliography{reference}

%
%
%

\end{document}